\documentclass[twocolumn,preprintnumbers,amsmath,amssymb,prl]{revtex4}
\usepackage[dvipdfmx]{graphicx}

\usepackage{graphicx}
\usepackage{dcolumn}
\usepackage{bm}
\usepackage{color}
\usepackage{rotating}
\usepackage{ulem} 

\begin{document}

\title{Diagonal Nematicity in the Pseudogap Phase of HgBa$_2$CuO$_{4+\delta}$ }

\author{H.\,Murayama$^1$}
\author{Y.\,Sato$^1$}
\author{R.\,Kurihara$^1$}
\author{S.\,Kasahara$^1$}
\author{Y.\,Mizukami$^2$}
\author{Y.\,Kasahara$^1$}
\author{H.\,Uchiyama$^{3,4}$}
\author{A.\,Yamamoto$^5$}
\author{E.-G.\,Moon$^6$}
\author{J.\,Cai$^7$} 
\author{J.\,Freyermuth$^7$} 
\author{M.\,Greven$^7$}
\author{T.\,Shibauchi$^2$}
\author{Y.\,Matsuda$^1$}
 \email{matsuda@scphys.kyoto-u.ac.jp}

\affiliation{$^1$ Department of Physics, Kyoto University, Kyoto 606-8502 Japan}
\affiliation{$^2$ Department of Advanced Materials Science, University of Tokyo, Chiba 277-8561, Japan} 
\affiliation{$^3$Materials Dynamics Laboratory, RIKEN SPring-8 Center, 1-1-1 Kouto, Sayo, Hyogo 679-5148, Japan}
\affiliation{$^4$Research and Utilization Division, Japan Synchrotron Radiation Research
Institute (SPring-8/JASRI), 1-1-1 Kouto, Sayo, Hyogo 679-5198, Japan}
\affiliation{$^5$  Graduate School of Engineering and Science, Shibaura Institute of Technology, 3-7-5 Toyosu, Koto-ku, Tokyo, 135-8584, Japan, }
\affiliation{$^6$ Department of Physics, Korea Advanced Institute of Science and Technology, Daejeon 305-701, Korea}
\affiliation{$^7$School of Physics and Astronomy, University of Minnesota, Minneapolis, Minnesota 55455, USA}



\maketitle

{\bf The pseudogap phenomenon in cuprates is the most mysterious puzzle in the research of high-temperature superconductivity~\cite{Keimer15}. In particular, whether the pseudogap is associated with a crossover or phase transition has been a long-standing controversial issue. 
The tetragonal cuprate HgBa$_2$CuO$_{4+\delta}$, with only one CuO$_2$ layer per primitive cell, is an ideal system  to tackle this puzzle. Here, we measure the anisotropy of magnetic susceptibility within the CuO$_2$ plane with exceptionally high-precision magnetic torque experiments. Our key finding is that a distinct two-fold in-plane anisotropy sets in below the pseudogap temperature $T^*$, which provides thermodynamic evidence for a nematic phase transition with broken four-fold symmetry. Most surprisingly, the nematic director orients along the diagonal direction of the CuO$_2$ square lattice, in sharp contrast to the bond nematicity reported in various iron-based superconductors~\cite{Fernandes14} and double-layer YBa$_2$Cu$_3$O$_{6+\delta}$~\cite{Daou10,Cyr-Choiniere15,Sato17}, where the anisotropy axis is along the Fe-Fe and Cu-O-Cu directions, respectively.  Another remarkable feature is that the enhancement of the diagonal nematicity with decreasing temperature is suppressed around the temperature at which short-range charge-density-wave (CDW) formation occurs~\cite{Tabis14, Tabis17}.  This is in stark contrast to YBa$_2$Cu$_3$O$_{6+\delta}$, where the bond nematicity is not influenced by the CDW.  Our result suggests a competing relationship between diagonal nematic and CDW order in  HgBa$_2$CuO$_{4+\delta}$.}


\begin{figure}[b]
	\vspace{-3cm}
	\begin{center}
		\includegraphics[width=1.5\linewidth]{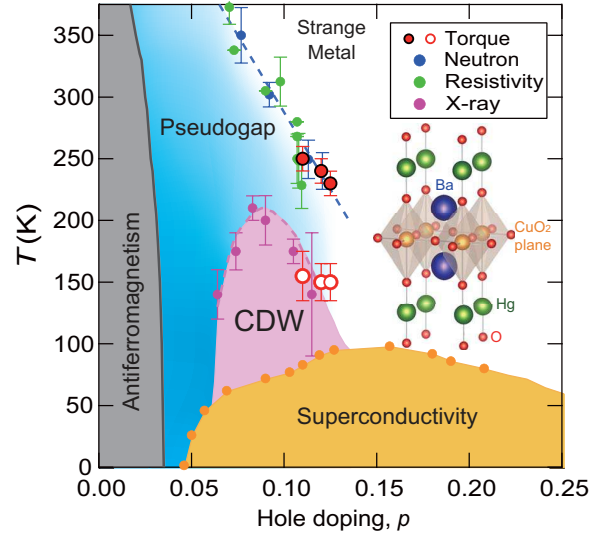}
	\caption{
		{\bf Temperature-doping phase diagram of Hg1201.}  Orange filled circles represent $T_c$ (see Methods).  Blue and green filled circles represent the pseudogap onset temperature $T^*$ determined by neutron scattering~\cite{Li08, Li11} and resistivity measurements~\cite{Barisic13}, respectively. Purple filled circles show CDW onset temperature $T_{\rm CDW}$ determined by resonant X-ray diffractions~\cite{Tabis14, Tabis17}.  Red filled circles represent the onset temperature of diagonal nematicity determined by in-plane torque magnetometry, which lies on the pseudogap line (blue dashed line).   Red open circles represent the temperature at which suppression of the nematicity occurs, which is close to  $T_{\rm CDW}$. As the antiferromagnetic order has not been reported in Hg1201,  it is shown in analogy with phase diagram of YBCO. The inset shows the crystal structure of Hg1201.	
	}
	\label{fig:figure1}
	\end{center}
\end{figure}

In hole-doped high transition temperature ($T_c$) cuprates,  anomalous electronic states, including Fermi arc, CDW and $d$-wave superconductivity, emerge below the pseudogap onset temperature $T^*$~\cite{Keimer15}.  The pseudogap formation has been discussed in terms of either a crossover phenomenon or a continuous phase transition.   In the former scenario, the pseudogap represents a precursory gap of the $d$-wave superconductivity, and the phase fluctuations of preformed Cooper pairs destroy the superconducting order.  In the latter scenario, the pseudogap emerges as a consequence of a spontaneous symmetry breaking. 
A continuous phase transition at $T^*$ is often argued to imply the presence of a quantum critical point, with associated fluctuations that may give rise to the high-$T_c$ superconductivity and strange-metal behaviours. Furthermore, the pseudogap order is thought to be intertwined (or compete) with other types of order, such as CDW order.  Until now, several types of broken symmetry, including broken translational, rotational, inversion and time reversal symmetry, have been deduced from various experiments, including scanning tunnelling microscopy~\cite{Kohsaka07,Lawler10, Parker10,Fujita14,Hamidian16}, polarized neutron scattering~\cite{Fauque06, Mangin-Thro14, Mangin-Thro17, Li08, Li11}, polar Kerr~\cite{Xia08}, optical~\cite{Zhao17}, and thermoelectric measurements~\cite{Daou10,Cyr-Choiniere15}.  Despite these tremendous efforts, the presence or absence of a continuous phase transition has been a highly controversial issue.

Recent torque-magnetometry of the anisotropic susceptibility within the $ab$ planes of YBa$_2$Cu$_3$O$_{6+\delta}$ (YBCO) revealed that the in-plane anisotropy displays a significant increase with a distinct cusp at $T^*$, consistent with the possible existence of a nematic phase transition~\cite{Sato17}. However, in YBCO, the four-fold ($C_4$) rotational symmetry is already broken due to the orthorhombic crystal structure with one-dimensional (1D) CuO chains, and thus no further rotational symmetry breaking is expected. Moreover, in bilayer YBCO, the coupling of the CuO$_2$ planes in the unit cell may lead to an additional effect on the symmetry breaking~\cite{Mangin-Thro17}. Therefore, the investigation of an underdoped single-layer system with tetragonal symmetry, such as hole-doped HgBa$_2$CuO$_{4+\delta}$ (Hg1201), is essential to clarify whether a nematic phase transition is an intrinsic and universal property of the high-$T_c$ cuprates. 
Figure\,1 displays the temperature-doping phase diagram of Hg1201. Similar to other hole-doped cuprates, the phase diagram contains antiferromagnetism, CDW, superconductivity and pseudogap regimes~\cite{Tabis14,Tabis17, Li08, Li11, Barisic13, Chan14, Ghiringhelli12, Chang12, Hucker14,Blanco-Canosa14}. 
The short-range CDW order, with the biaxial wave vectors close to 0.28 r.l.u., forms a dome-shaped boundary inside the pseudogap regime~\cite{Tabis14,Tabis17}. 
In zero magnetic field, CDW domains with typical size of a few nanometers appear.
Within the domains, both the $C_4$ and translational symmetries are strongly broken~\cite{Fujita14,Hamidian16}.

\begin{figure}[hbtp]
	\begin{center}
		\includegraphics[width=0.8\linewidth]{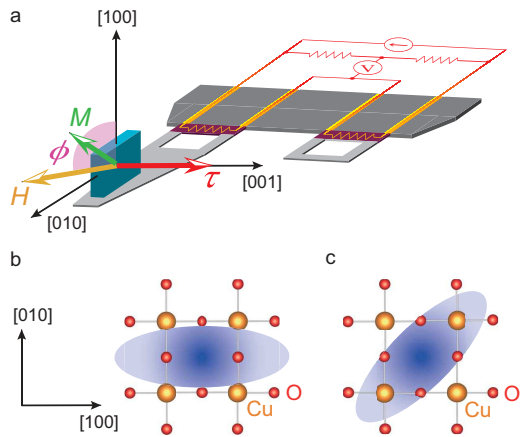}
	\caption{{\bf Torque magnetometry determination of nematic order.}
		{\bf a},  The experimental configuration for in-plane  torque magnetometry.  The magnetic field is rotated within the tetragonal $ab$ plane.    A single-crystalline sample of Hg1201 is mounted on the piezo-resistive lever which forms an electrical bridge circuit with the neighbouring reference lever. {\bf b}, Schematic picture of bond nematicity with $B_{1g}$ symmetry, where the nematicity appears along the Cu-O-Cu direction.  For this nematicity, $\chi_{aa}\neq \chi_{bb}$ and $\chi_{ab}=0$.   {\bf c}, Diagonal nematicity with $B_{2g}$ symmetry, where the nematic director is along the diagonal direction of CuO$_2$ square lattice.	 For this nematicity, $\chi_{aa}=\chi_{bb}$ and $\chi_{ab}\neq 0$.}
	\label{fig:figure2}
	\end{center}
\end{figure}

\begin{figure}[hbtp]
	\begin{center}
		\includegraphics[width=1.0\linewidth]{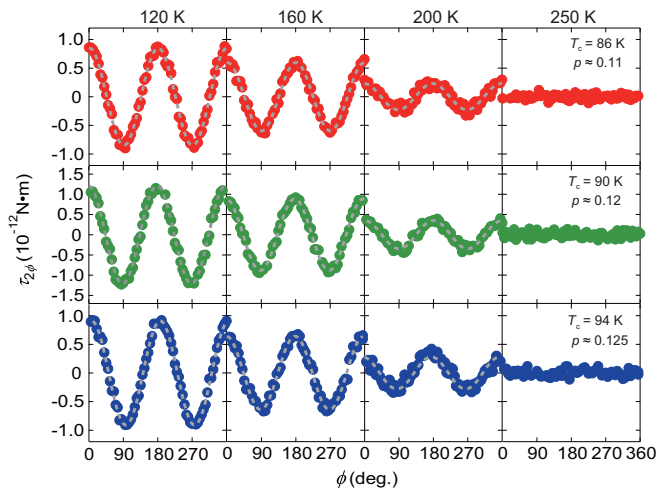}
	\caption{{\bf Two-fold oscillations of magnetic torque in the CuO$_2$ planes.}
		 Upper panels show the torque curves $\tau_{2\phi}$ as a function of the azimuthal angle $\phi$ for $p\approx 0.11$. Middle and lower panels show $\tau_{2\phi}$ for $p\approx0.12$ and 0.125, respectively. 
	}
	\label{fig:figure3}
	\end{center}
\end{figure}

\begin{figure*}[t]   
\begin{center}
		\includegraphics[width=1.0\linewidth]{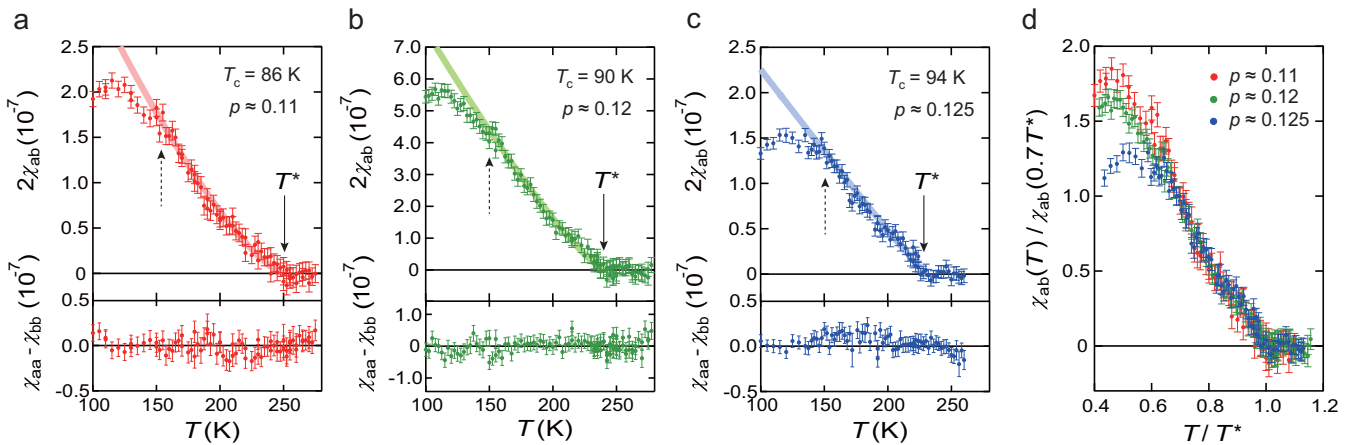}
	\caption{{\bf Anisotropy of magnetic susceptibility.}
		{\bf a}, Temperature dependence of $2\chi_{ab}$ (upper panel) and $\chi_{aa}-\chi_{bb}$ (lower panel) for $p\approx$ 0.11.     {\bf b},{\bf c}, Same plots for $p\approx$ 0.12 and  0.125, respectively.   Solid arrows indicate the onset temperatures of $\chi_{ab}$, which well coincides with $T^*$, as shown in Fig.\,1.  Dashed arrows indicate the temperatures at which $\chi_{ab}$ deviates from the extrapolation from high temperature shown by the bold lines. These temperatures are close to $T_{\rm CDW}$, as shown in Fig.\,1. 
	}
	\label{fig:figure4}
	\end{center}
\end{figure*}
	
The measurement of the magnetic torque  $\bm{\tau} = \mu_0 V{\bm M} \times {\bm H}$ has a high sensitivity for the detection of  magnetic anisotropy, where  $\mu_{0}$ is space permeability, $V$ is the sample volume, and  $\bm{M}$ is the magnetization induced by external magnetic field $\bm{H}$ (Fig.\,2a)~\cite{Okazaki11,Kasahara12}. Torque is a thermodynamic quantity which is a differential of the free energy with respect to angular displacement. Torque measurements performed for a range of directions of $\bm{H}$ within the tetragonal $ab$ plane in Hg1201 provide a stringent test of whether or not the pseudogap state breaks the crystal four-fold symmetry. In such a geometry, $\bm{\tau}$ is a periodic function of double the azimuthal angle $\phi$ measured from the $a$ axis:
\begin{equation}
\tau_{2\phi}=\frac{1}{2}\mu_{0}H^{2}V[(\chi_{aa}-\chi_{bb})\sin2\phi-2\chi_{ab}\cos2\phi]
\end{equation}
where  the susceptibility tensor $\chi_{ij}$ is defined by $M_{i}=\sum_{j}\chi_{ij}H_{j}$ $(i, j=a, b, c)$. In a system with tetragonal symmetry, $\tau_{2\phi}$ should be zero because $\chi_{aa}=\chi_{bb}$ and $\chi_{ab}=0$.  Non-zero values of $\tau_{2\phi}$  appear when a new electronic or magnetic state emerges that breaks the tetragonal symmetry; $C_4$  rotational symmetry breaking is revealed by $\chi_{aa}\neq\chi_{bb}$ and/or $\chi_{ab}\neq 0$. The former and the latter states are illustrated in Figs.\,2b and c, where the $C_4$ symmetry breaking occurs along [100]/[010] direction (bond nematicity with $B_{\rm 1g}$-symmetry) and [110] direction (diagonal nematicity with $B_{\rm 2g}$-symmetry) of the CuO$_2$ plane.

The upper, middle and lower panels of Fig.\,3 depict the magnetic torque curves measured as a function of $\phi$ for Hg1201 crystals with three different hole concentrations $p\approx$ 0.11, 0.12, and 0.125 with $T_c = $ 86, 90 and 94\,K, respectively.  The approximate crystal sizes are $150 \times 140 \times 30$ $\mu$m$^3$ ($p \approx 0.11$),  $90 \times  90 \times 40$  $\mu$m$^3$ ($p \approx 0.12$), and $120 \times  150 \times 50$  $\mu$m$^3$ ($p \approx 0.125$).  To exclude two-fold oscillations appearing as a result of misalignment, the magnetic field ($|\mu_0\bm{H}|$ = 4\,T) is precisely applied in the $ab$ plane with out-of-plane misalignment less than 0.1$^\circ$  by controlling two superconducting magnets and a rotating stage (see Figs.\,S1 and S2 and Method). For all crystals the two-fold oscillation is absent at high temperatures, which is consistent with the tetragonal crystal symmetry. At low temperatures, however, the emergence of distinct two-fold oscillations $\tau_{2\phi}$ is observed. This provides direct evidence for nematicity, indicating that the four-fold rotational symmetry at high temperature is broken down to the  two-fold rotational symmetry ($C_4 \rightarrow C_2$).  Moreover, the two-fold oscillation follows the functional form $\tau_{2\phi}=A_{2\phi}\cos 2\phi$, i.e.,$\chi_{ab}\neq 0$ and $\chi_{aa}=\chi_{bb}$, which demonstrates the emergence of diagonal nematicity.

Figures\,4a, b and c depict the $T$-dependence of 2$\chi_{ab}$ along with that of $\chi_{aa}-\chi_{bb}$ for $p\approx$ 0.11, 0.12 and 0.125, respectively. While $\chi_{aa}-\chi_{bb}$ is almost temperature-independent and negligibly small within the resolution, $2\chi_{ab}(T)$ shows characteristic temperature dependence.  For all doping levels,  $\chi_{ab}=0$  at high temperatures. At the temperatures shown by solid arrows in Figs.\,4a,\,b, and c, $\chi_{ab}$ becomes finite and grows rapidly as the temperature is lowered. We plot these temperatures by red filled circles in Fig.\,1. Obviously, the onset temperatures of the nematicity lie on the pseudogap line in the doping-temperature phase diagram determined by other methods, consistent with spontaneous macroscopic $C_4$ rotational symmetry breaking at $T^*$, i.e., with the notion that the onset of the pseudogap is characterized by a nematic phase transition.

As a  natural consequence of the tetragonal crystal structure,  the pseudogap  phase with $C_2$ symmetry forms domains with different preferred directions in the $ab$ plane.   We note that although the magnitude of $\chi_{ab}$ for the three crystals is of the same order, $2\chi_{ab}$ for $p\approx$ 0.12 is a few times larger than for $p\approx$ 0.11 and 0.125.   As the crystal volume of $p\approx$\,0.12 is  the smallest among the three crystals, a domain size on the order of tens of micrometers, which gives rise to unequal numbers of domains in the crystals, would explain the present results. The torque curves remain unchanged for field-cooling conditions at different field angles.  To check the influence of the strain on the side of the crystal attached to the cantilever, we measured the torque after remounting the crystal rotated by 90$^\circ$ (Figs.\,S3a and S3b).  The direction of the nematicity is unchanged relative to the crystal axes after the crystal rotation.   Although the magnitude of $\chi_{ab}$ is enhanced, possibly due to a change in the imbalance of the number of the domains, temperature dependence of $\chi_{ab}(T)$ is essentially the same in a wide temperature range below $T^*$ (Fig.\,S3c).  This implies that a large fraction of the domains are pinned by the  underlying crystal conditions, such as internal stress, disorder, and crystal shape, which may be consistent with the absence of hysteresis and thermal history effects.  It also suggests that uniaxial pressure may reverse the direction of the nematicity.  We note that a somewhat analogous situation is encountered in polar Kerr-effect measurements on YBCO, where the sign of the time-reversal symmetry breaking is fixed at temperatures significantly above $T^*$  \cite{Xia08}.

In Fig.\,4d, the scaled values $\chi_{ab}(T) / \chi_{ab}(0.7T^*) $ for  $p\approx$ 0.11, 0.12 and 0.125 are presented.  The three curves collapse onto the universal curve within the error bars down to $T\sim0.65T^*$.  The scaling behaviour as well as characteristic super-linear temperature dependence has also been reported in YBCO \cite{Sato17} and might therefore be universal properties of the nematic transition.

Deep inside the nematic phase, at temperatures well below $T^*$, $\chi_{ab}(T)$ exhibits another anomaly. For $p \approx 0.125$,  $\chi_{ab}(T)$ is strongly suppressed below $T\sim150$\,K (Fig.\,4c).  Although such an anomaly is less pronounced for $p=0.11$ and 0.12,  $\chi_{ab}(T)$ shows a deviation from an extrapolation from higher temperatures at $T=$ 140-160\,K, as shown by red and green lines in Figs.\,4a and b. The observed  suppression of the nematicity well below $T^*$ is in stark contrast to YBCO, in which the nematicity grows monotonically with decreasing temperature without discernible anomaly down to $T_c$.   These temperatures are close to the CDW transition temperature $T_{\rm CDW}$ determined by resonant X-ray diffraction measurements~\cite{Tabis14,Tabis17}.   It has been shown that below this temperature,  the Hall constant levels off and the resistivity shows $T^2$ dependence~\cite{Barisic13}.  The CDW modulation is along [100]/[010], which is the same direction as the bond nematicity in YBCO, but differs from the diagonal nematicity in Hg1201. The fact that $\chi_{aa}-\chi_{bb}(T)$ remains negligibly small below $T_{\rm CDW}$ implies that the CDW domain size is much smaller than the sample dimensions, which results in the cancellation of the $\sin 2\phi$ oscillations with opposite signs from different domains. This is consistent with the scattering experiments which revealed CDW correlations of a few nm~\cite{Tabis14,Tabis17}. 

It should be stressed that the nature of the nematicity in Hg1201 differs from that in YBCO in several regards.  First, purely spontaneous $C_4\rightarrow C_2$ rotational symmetry breaking of electron system occurs in Hg1201 with tetragonal crystal symmetry.  Second, the diagonal nematicity with $B_{2g}$ symmetry in Hg1201 is in stark contrast to the bond nematicity with $B_{1g}$ symmetry in YBCO.  Third, while the growth of the diagonal nematicity is suppressed by the CDW formation in Hg1201, no discernible anomaly is observed at $T_{\rm CDW}$ in YBCO.


Our results shed new light on the pseudogap physics. Nematicity in the pseudogap phase is universal regardless of the number of CuO$_2$ layers per primitive cell. Moreover, the diagonal nematicity in Hg1201, observed in this study for the first time, is associated with the $B_{\rm 2g}$ representation, in sharp contrast to the CDW order along the bond direction ($B_{\rm 1g}$), which demonstrates that the nematicity is not a precursor of the CDW.  Notice that in other cuprates such as YBCO and Bi$_2$Sr$_2$CaCu$_2$O$_{8+\delta}$,  nematicity and CDW develop along the same bond direction and the precursor issue has not been clear.

The super-linear temperature behaviour of the nematicity in YBCO and Hg1201 suggests that pseudogap transition may not belong to the two- or three-dimensional Ising universality class.  One possible explanation is that the nematicity may not be a primary order parameter, whereas another possibility is that pseudogap phenomena are associated with physics beyond the Landau paradigm. 

Next, we discuss possible explanations of the diagonal nematicity.  Our results suggest that the diagonal nematicity is unlikely due to a Pomeranchuk instability of the Fermi surface, which prefers symmetry breaking along the bond direction ~\cite{Halboth00}. 
We note that we cannot rule out the possibility of complete rotational symmetry breaking ($C_4 \rightarrow C_1$).  The pattern of an intra-unit-cell loop-current order with $C_1$ symmetry may be consistent with the diagonal direction~\cite{Varma06}. 
However, the interpretation of the polarized neutron results \cite{Fauque06,Mangin-Thro14,Mangin-Thro17,Li08,Li11} in terms of loop-current-order is not unique, and the observed magnetism appears to be dynamic rather than truly static.
In addition, it is an interesting open question whether the loop current order would explain the bond nematicity in 
bilayer cuprates such as YBCO.  
Recent resistivity measurements on tetragonal La$_{2-x}$Sr$_x$CuO$_4$ thin films report broken $C_4$ symmetry even above $T^*$.  Moreover the direction of the nematicity is not fixed by the crystal axes, and depends on temperature and hole concentration.  
At the present stage, the relationship between this transport work and our thermodynamic result is an open question~\cite{Wu17}.

More scenarios have been suggested recently. One is based on octupolar order of Cu-3$d$ orbital for bond nematicity under the Landau paradigm~\cite{Hitomi16}, another is based on percolation of local pseudogaps which may be consistent with nematic domain phenomena near $T^*$~\cite{Pelc17}, and a third is based on phenomena beyond the Landau paradigm such as doped quantum spin liquid, which may explain the super-linear onset intrinsically~\cite{Lee18}. 
Further theoretical and experimental investigations are highly desired to pin down a mechanism of the diagonal nematicity.  

\bigskip 

\noindent
{\bf Acknowledgements}\\
We thank A. Fujimori, T. Hanaguri, S. Kivelson, H. Kontani, P.A. Lee, D. Pelc, S. Sachdev, L. Taillefer, T. Tohyama, C. Varma, H. Yamase,  Y. Yanase, and G. Yu for fruitful discussions. This work was supported by Grants-in-Aid for Scientific Research (KAKENHI) (Nos. 25220710, 15H02106, 15H03688, 16K13837) and on Innovative Areas ``Topological Material Science" (No. 15H05852) from Japan Society for the Promotion of Science (JSPS).  This work was partly performed using facilities of the Institute for Solid State Physics, the University of Tokyo. H.U. thanks Alfred Baron for making the work possible in Materials Dynamics laboratory, RIKEN SPring-8 Center.   The work at the University of Minnesota was funded by the Department of Energy through the University of Minnesota Center for Quantum Materials under DE-SC-0016371.  E.-G.M. acknowledges the financial supports from the POSCO Science Fellowship of POSCO TJ Park Foundation and NRF of Korea under Grant No. 2017R1C1B2009176. 

\bigskip

\noindent
{\bf Author contributions}\\
H.U., A.Y., J.C., J.F. and M.G. grew the high-quality single-crystalline samples. H.M., Y.S.,R.K. and S.K. performed the magnetic torque measurements. H.M. and Y. Mizukami performed the X-ray diffraction measurements.  H.M., Y.S., S.K., Y.K., E.-G.M. and Y. Matsuda analysed the data.   H.M., S.K., E.-G.M.,  T.S., M.G. and Y. Matsuda discussed and interpreted the results  and prepared the manuscript.

\newpage
\noindent {\bf Methods}\\
\noindent {\bf Materials.}
We have studied single crystals of Hg1201 grown by two different techniques. For $p \approx 0.11$ and 0.125, small single crystals were grown by a solid-state reaction method~\cite{Yamamoto01, Yamamoto00}. For $p \approx 0.12$, a small crystal was cut from   a large crystal grown by a flux method~\cite{Zhao06}.  The hole concentration for $p \approx 0.12$ was controlled by annealing the crystal at high temperatures under oxygen atmosphere. The hole doping levels were determined from the superconducting transition temperature $T_c$ by magnetization measurements~\cite{Yamamoto00}. The crystals exhibit sharp superconducting transitions with   $T_c$  of 86, 90 and 94\,K for $p \approx 0.11, 0.12$ and 0.125, respectively. The directions of the crystalline axes were determined by X-ray diffraction measurements.  

\bigskip

\noindent {\bf Torque magnetometry.}
Magnetic torque was measured by the piezo-resistive micro-cantilever technique~\cite{Sato17, Okazaki11, Kasahara12}. Tiny single crystals were carefully mounted onto the piezo-resistive lever which forms an electrical bridge circuit with the neighbouring reference lever.

For the precise measurements of the in-plane magnetic torque, we use a system consisting of a 2D vector magnet and a mechanical rotator (Fig.\,S1), which enables us to rotate the magnetic field $\bm{H}$  within the $ab$ plane.   
Small out-of-plane misalignment which is inevitable when mounting the sample is precisely determined by measuring the out-of-plane torque $\tau$ as a function of the angle from $z$ axis (Fig.\,S1).  Figure\,S2a shows typical curves of $\tau$ for $p\approx 0.11$ at 80 K in the superconducting state.   When $\bm{H}$ is rotated within the $ac$ plane,  $\tau$ exhibits  a very sharp change at ${\bm H}\parallel ab$ plane.  Figure\,S2b represents the out-of-plane misalignment $\Delta \theta_m$ as a function of azimuthal angle $\phi$ for $p\approx 0.11$.    Computer controlling the vector field and mechanical rotator systems, we eliminate the misalignment and rotate $\bm{H}$ within the $ab$ plane with the accuracy better than 0.1$^\circ$.

Figures\,S4a and b shows typical out-of-plane anisotropy $\Delta \chi_{\perp}=\chi_{cc}-\chi_{\parallel}$  in the normal state ($T > T_c$)  for $p\approx0.11$ and 0.125, respectively, obtained  by rotating $\bm{H}$ within a plane including the $c$ axis. The $\tau(\theta)$ curves  exhibit purely paramagnetic response with no discernible hysteresis components, 
and are well fitted by 
\begin{equation}
\tau_{2\theta}(\theta, T,H)=\frac{1}{2}\mu_0H^2V\Delta\chi_{\perp}\sin 2\theta, 
\end{equation}
which yields $\pi$ periodic oscillations with respect to the $\theta$ rotation.  Here $\theta$ is the polar angle from the $c$ axis,  $\Delta\chi_\perp = \chi_{cc} - \chi_{\parallel}$ is the difference between the $c$ axis and the in-plane susceptibilities, and $\chi_{\parallel}$ is the average of in-plane susceptibilities. 
From the amplitude of the $\tau(\theta)$ curves, the temperature dependences of $\Delta\chi_\perp$ are obtained.  At high temperatures, the magnitude of $\Delta\chi_\perp$ changes nearly linearly with temperature, while the data deviate downward from the $T$-linear behaviour below $T^*$, which represents the pseudogap formation~\cite{Sato17}.

 
\bigskip

\begin{figure}[h]
	\includegraphics[width=\linewidth]{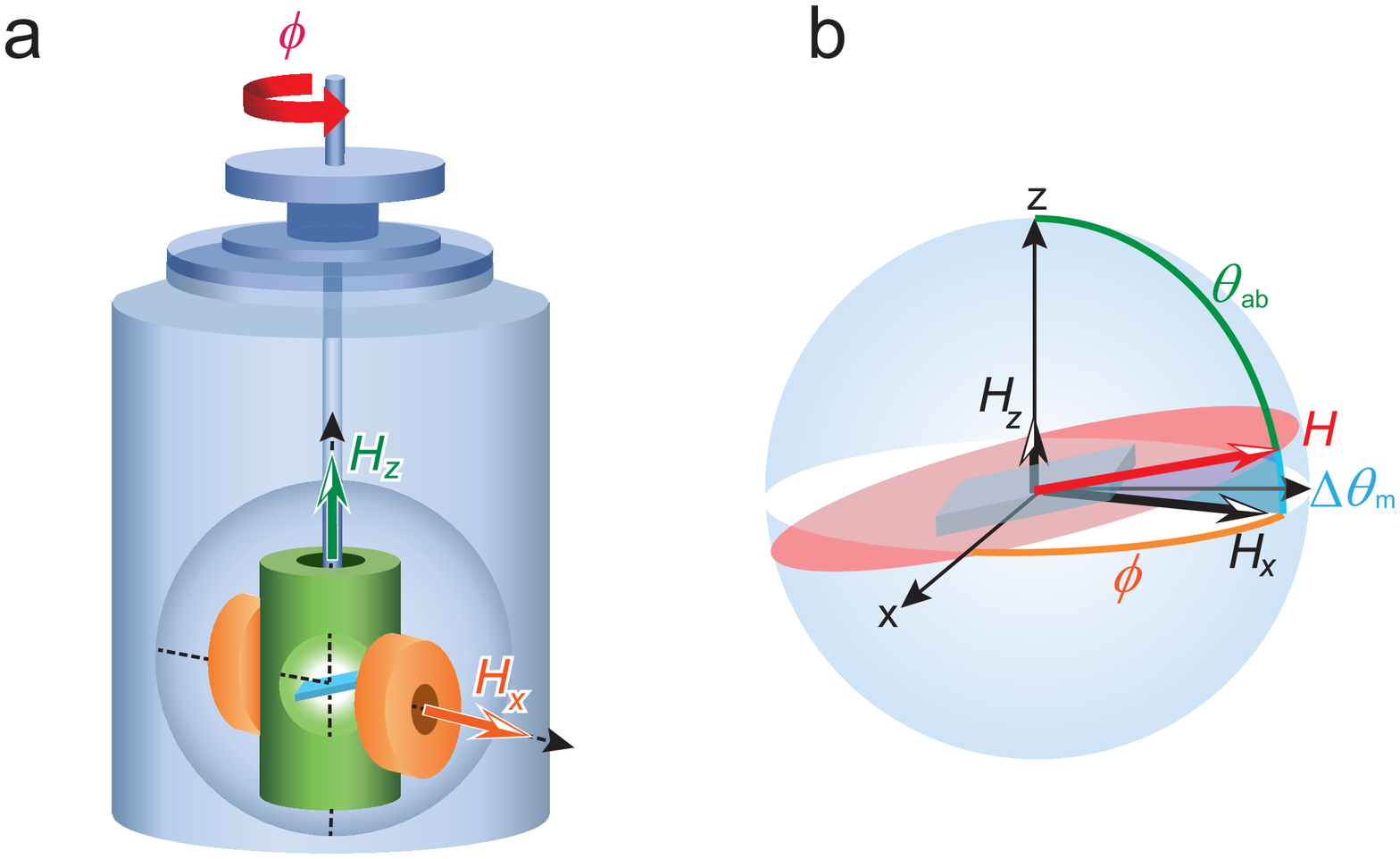}
\end{figure}

\noindent 
{\bf  Figure S1 $|$ Schematic figures of the torque measurement system.}
{\bf a,} Schematics of experimental apparatus. To apply magnetic field {\boldmath $H$} with high accuracy relative to the crystal axes,  we used a system with two superconducting magnets generating {\boldmath $H$} in two mutually orthogonal directions ($x$ and $z$) and a mechanical rotating stage at the top of the Dewar.  {\bf b,} Schematics of in-plane $\phi$-scan measurements. {\boldmath $H$}  was applied in the $ab$ plane with high alignment precision (misalignment of less than 0.1$^{\circ}$). $\theta_{ab}$ is the  angle between the $z$ axis and the $ab$ plane.  $\Delta \theta_m$ is the misalignment from $H_x$ (see Fig.\,S2).


\begin{figure}[h]
	\includegraphics[width=\linewidth]{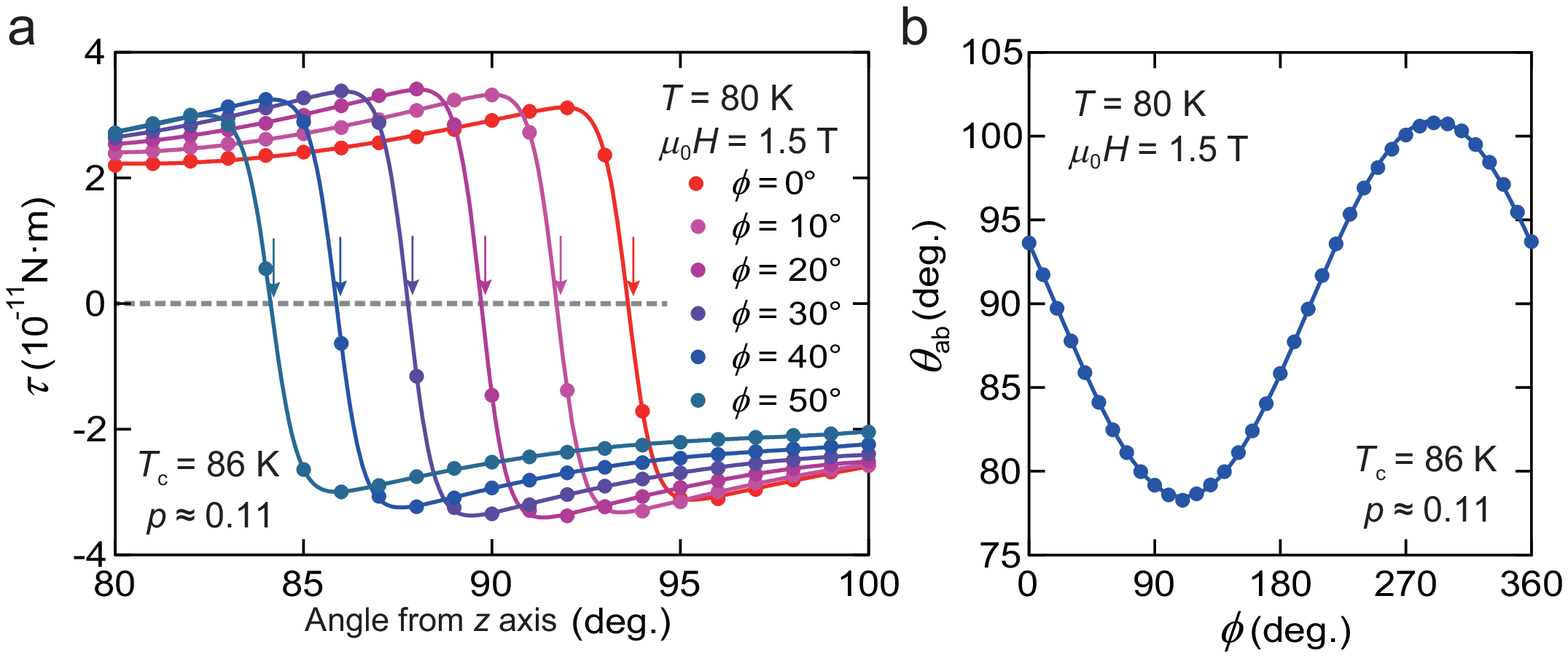}
\end{figure}

\noindent 
{\bf Figure S2 $|$ Determination of the misalignments in Hg1201.}  
{\bf a}, Torque $\tau$ plotted as a function of angle from $z$ axis for $p \approx 0.11$ in the superconducting state at $T = 80$ K in magnetic field of $\mu_0H=1.5$\,T rotated across the $ab$ plane at several $\phi$.  The torque curve is completely reversible as a function of  angle at this temperature.   $\tau$ abruptly changes its sign when crossing the $ab$ plane. The arrows indicate the angle $\theta_{ab}$, at which $\tau$ vanishes.   {\bf b},  $\theta_{ab}$  plotted as a function of  $\phi$.  The misalignment  of the applied field from the $ab$ plane is given as $\Delta \theta_m=\theta_{ab}-90^{\circ}$.   $\Delta \theta_m$ is nearly perfectly sinusoidal as a function of $\phi$.   This misalignment is   eliminated by computer controlling the vector magnet and mechanical rotator shown in Fig.\,S1.

\begin{figure}[h]
	\includegraphics[width=\linewidth]{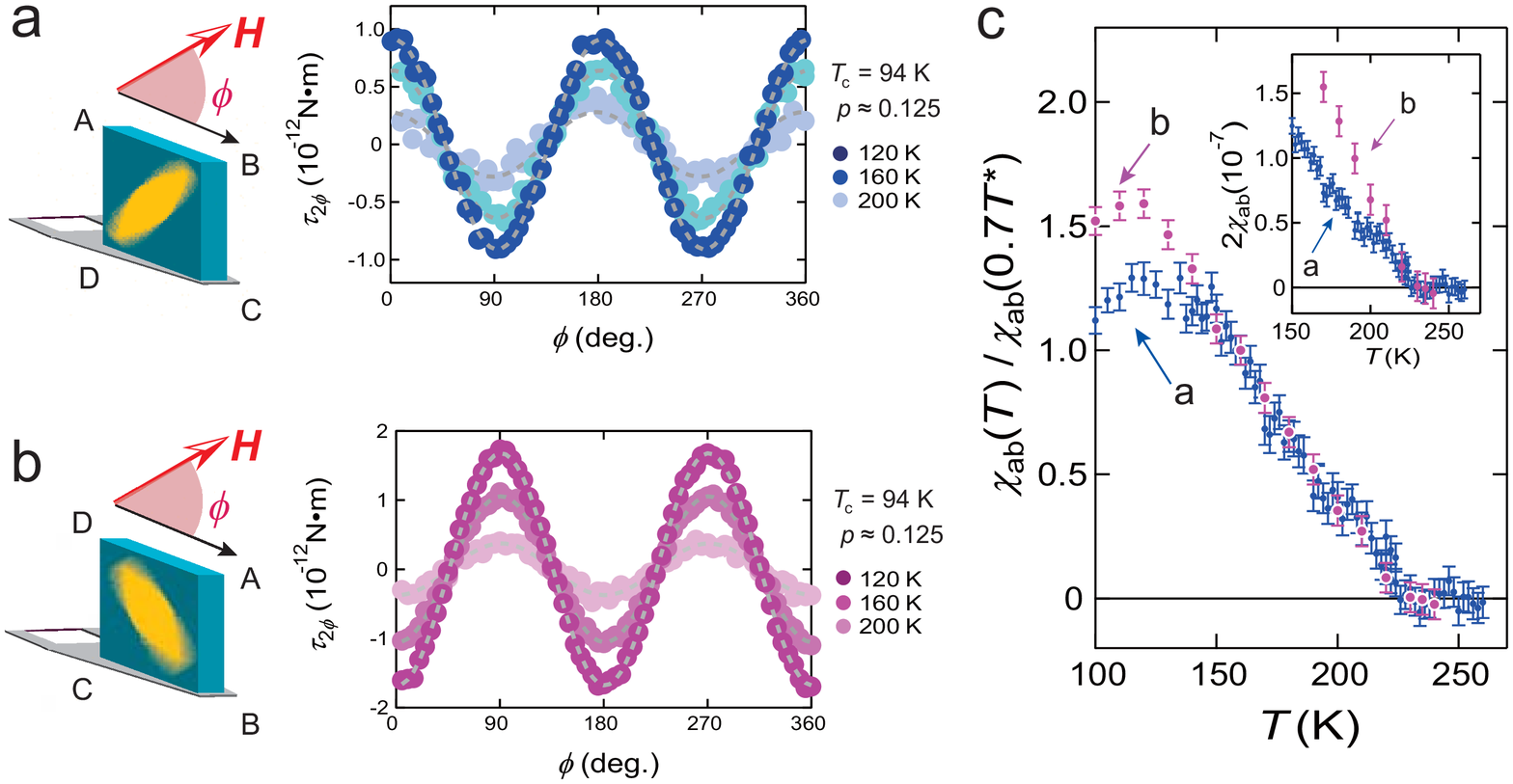}
\end{figure}

\noindent 
{\bf Figure S3 $|$ The influence of strain on the side of the crystal attached to the cantilever.} 
{\bf a},{\bf b,}  (left) Schematic of torque measurements in two different configurations.  Yellow ellipses indicate the nematic directions.   In {\bf b},  the torque is measured after remounting the crystal rotated by 90 degrees relative to the configuration illustrated in {\bf a}.    Right figures depict  $\tau_{2\phi}$ vs. $\phi$ for the configurations, respectively, for the crystal with $p$=0.125.   The direction of the nematicity is unchanged relative to the crystal axes after the crystal  rotation.  {\bf c,}  Inset depicts temperature dependence of $\chi_{ab}$ for configurations shown in {\bf a} and {\bf b}.     Main panel depicts $\chi_{ab}(T)/\chi_{ab}(0.7T^*)$ for both configurations.  Temperature dependence  is essentially the same down to 150\,K below which CDW is formed.

\begin{figure}[h]
	\includegraphics[width=\linewidth]{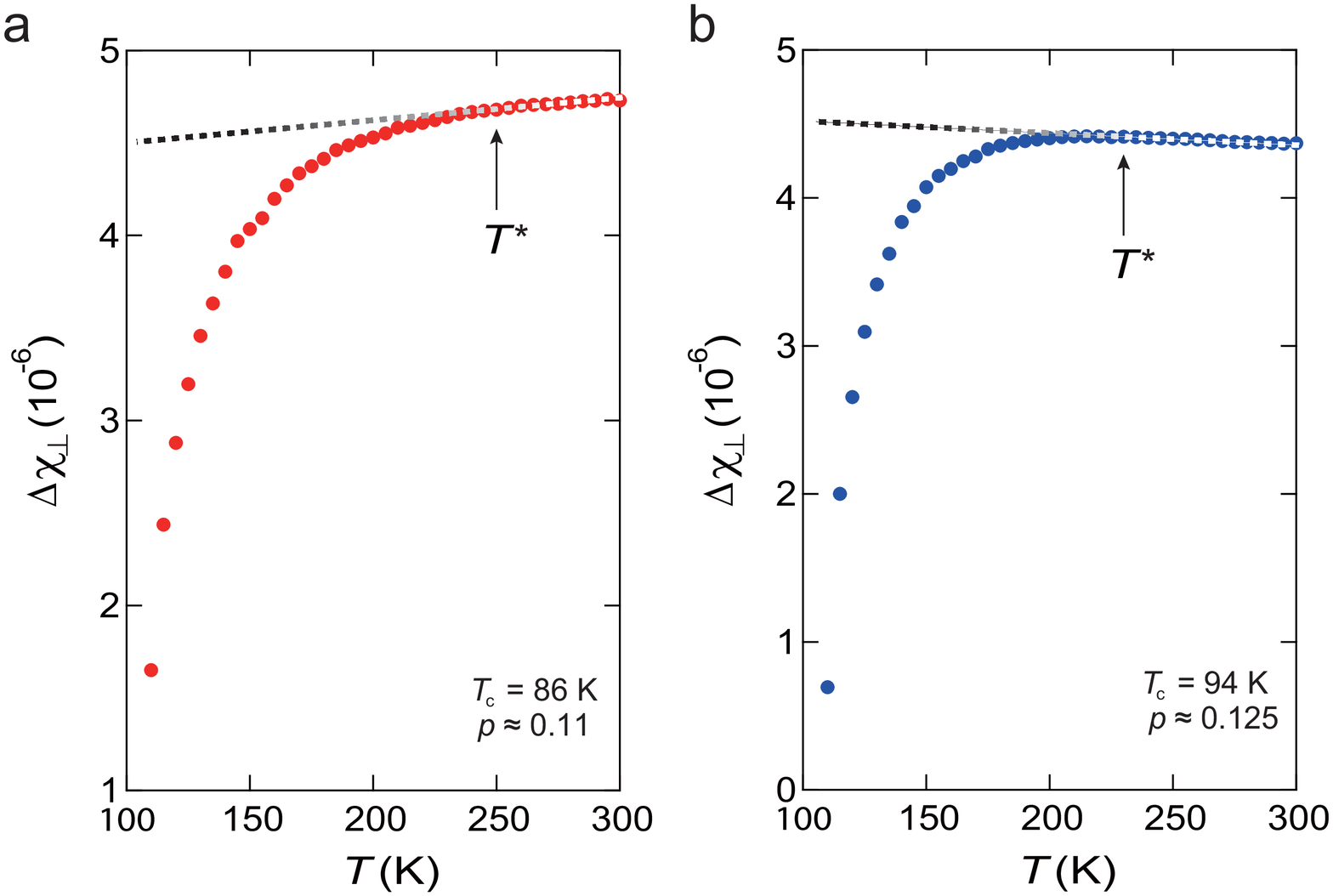}
\end{figure}

\noindent 
{\bf Figure S4 $|$ Out-of-plane anisotropy of the magnetic susceptibility in Hg1201.} Temperature dependence of the out-of-plane anisotropy of the magnetic susceptibility, $\Delta\chi_\perp\equiv \chi_{cc}-\chi_{\parallel}$, determined from the $\tau(\theta)$ curves for $p\approx$ 0.11 ({\bf a}) and 0.125 ({\bf b}). Dashed lines are $T$-linear fits at high temperatures. Below $T^*$, $\Delta\chi_\perp$ shows deviations from the high-temperature linear behaviour.

\end{document}